\documentclass[12pt]{article}
\newcommand\mysection{\setcounter{equation}{0}\section}

\def\MSbar{\relax\ifmmode\overline{\rm MS}\else{$\overline{\rm MS}${ }}\fi}

\def\fun#1#2{\lower3.6pt\vbox{\baselineskip0pt\lineskip.9pt
  \ialign{$\mathsurround=0pt#1\hfil##\hfil$\crcr#2\crcr\sim\crcr}}}

\def\eV{{\rm e\kern-0.12em V}}            
  
\def\half{{\textstyle {1\over2}}}
  \def\quart{{\textstyle {1\over4}}}

\def \al {\relax\ifmmode{\alpha}\else{$\alpha${ }}\fi}
\def \be {\relax\ifmmode{\beta}\else{$\beta${ }}\fi}

\def\beq{\begin{equation}}   \def\eeq{\end{equation}}

\def \as{\relax\ifmmode\alpha_s\else{$\alpha_s${ }}\fi}
\def \pt{\relax\ifmmode{p_t}\else{$p_t${ }}\fi}

\def\eps{\relax\ifmmode\epsilon\else{$\epsilon${ }}\fi}

\def\ee{\relax\ifmmode{e^+e^-}\else{${e^+e^-}$}\fi}
\def\qq{\relax\ifmmode{q\overline{q}}\else{$q\overline{q}${ }}\fi}

\newskip\humongous \humongous=0pt plus 1000pt minus 1000pt
\def\caja{\mathsurround=0pt}
\def\eqalign#1{\,\vcenter{\openup1\jot
\caja   \ialign{\strut \hfil$\displaystyle{##}$&$
\displaystyle{{}##}$\hfil\crcr#1\crcr}}\,}

\newif\ifdtup

\def\eqal2#1{\,\vcenter{\openup1\jot
\caja   \ialign{\strut \hfil$\displaystyle{##}$&\hfil$
\displaystyle{{}##}$\hfil &$
\displaystyle{{}##}$\hfil\crcr#1\crcr}}\,}

\def\np#1#2#3{{\em Nucl.Phys.}~\underline{B#1} (19#3) #2}
\def\pl#1#2#3{{\em Phys.Lett.}~\underline{#1B} (19#3) #2}
\def\pr#1#2#3{{\em Phys.Rev.}~\underline{D#1} (19#3) #2}

\def\prl#1#2#3{{\em Phys.Rev.Lett.}~\underline{#1} (19#3) #2}

 \def\cite#1{[\ref{#1}]}
 \def\citd#1#2{[\ref{#1},\ref{#2}]}

 \def\citm#1#2{[\ref{#1}--\ref{#2}]}
\def\Apr{A^\prime}

\def\cF{{\cal{F}}}

\def\at{\al_{\mbox{\scriptsize eff}}}

\def\aPT{\as^{\mbox{\scriptsize PT}}}
\def\Freg{\cF^{\mbox{\scriptsize reg}}}
\def\CNP{C^{\mbox{\scriptsize NP}}}

\def\Li{\mbox{Li}_2}

\def\cFr{{\cal{F}}^{(r)}}
\def\cFv{{\cal{F}}^{(v)}}
\def\re#1{(\ref{#1})}

\begin{document}
\begin{titlepage}
\begin{flushright}
Cavendish-HEP-96/1\\
hep-ph/9604388
\end{flushright}              
\vspace*{\fill}
\begin{center}
{\Large \bf
Power Corrections and Renormalons in\\[1ex]
Deep Inelastic Structure Functions\footnote{Research supported in
part by the U.K. Particle Physics and Astronomy Research Council and by
the EC Programme ``Human Capital and Mobility", Network ``Physics at
High Energy Colliders", contract CHRX-CT93-0357 (DG 12 COMA).}}
\end{center}
\par \vskip 5mm
\begin{center}
        M.\ Dasgupta and B.R.\ Webber \\
        Cavendish Laboratory, University of Cambridge,\\
        Madingley Road, Cambridge CB3 0HE, U.K.
\end{center}
\par \vskip 2mm
\begin{center} {\large \bf Abstract} \end{center}
\begin{quote}
We study the power corrections (infrared renormalon contributions)
to the coefficient functions for non-singlet deep inelastic structure
functions due to gluon vacuum polarization insertions in one-loop graphs.
Remarkably, for all the structure functions $F_1$, $F_2$, $F_3$ and $g_1$,
there are only two such contributions, corresponding to $1/Q^2$ and
$1/Q^4$ power corrections.  We compute their dependence on Bjorken $x$.
The results could be used to model the dominant higher-twist contributions.
\end{quote}
\vspace*{\fill}
\begin{flushleft}
     Cavendish--HEP--96/1\\
     April 1996
\end{flushleft}
\end{titlepage}

\mysection{Introduction}

A number of recent papers have discussed power corrections
to non-singlet deep inelastic structure functions and sum rules,
using either the language of infrared renormalons \citm{BroKat}{stein}
or a dispersive approach based on assumed analyticity
properties of the running coupling \cite{BPY}.
The contributions computed correspond to the sum
of vacuum polarization insertions
(`renormalon chains' \cite{renormalons})
on the gluon line in one-loop corrections to the
coefficient function.
Phenomenologically, it appears \citd{stein}{BPY}
that the corrections computed in this way provide a 
good guide to the form of the higher-twist
contributions observed experimentally. This
may be understandable in terms of the notion of a
universal infrared-finite effective coupling \cite{BPY}.

In the present paper we point out that, within the
framework of the approaches mentioned, there are
only two power-behaved contributions,
corresponding to $1/Q^2$ and $1/Q^4$ corrections.
This appears to be a special feature of deep
inelastic structure functions. It is not the case
for $\ee$ fragmentation functions, for example,
where the same graphs generate an infinite set
of $1/Q^{2p}$ corrections. Of course, higher
power corrections might also be generated by
inserting renormalon chains in multi-loop
graphs. Nevertheless, to the extent that the
calculations based on one-loop graphs are
phenomenologically successful, one may regard
higher-power contributions as correction
terms in the deep inelastic case.

We present our calculation using the dispersive
approach introduced in Ref.~\cite{BPY}. This
makes it simple to compute the power-behaved
contributions and their dependence on Bjorken $x$.
As mentioned above, the results could be useful
as a model for the dominant higher-twist contributions.
They are expressed in terms of two non-perturbative
parameters, which could be determined experimentally
by fitting deep inelastic data at moderate $Q^2$.

\mysection{Dispersive method}

The method of Ref.~\cite{BPY} starts from a (formal) dispersion
relation for the QCD running coupling $\as(k^2)$ of the form
\beq\label{nonAb-disrel}
\as(k^2) = - \int_0^\infty \frac{d\mu^2}{\mu^2+k^2} \>\rho_s(\mu^2)
\,, \;\;\;\;\;\;\;
\rho_s(\mu^2) =
-\frac{1}{2\pi i}\, \mbox{Disc} \left\{\as(-\mu^2)\right\} \,.
\eeq
Introducing the effective coupling $\at(\mu^2)$, defined in
terms of the `spectral function' $\rho_s(\mu^2)$ by
\beq\label{aeffrel}
 \rho_s(\mu^2) =\frac{d}{d\ln\mu^2} \at(\mu^2) \>,
\eeq
it follows that
\beq\label{atint}
\as(k^2) = k^2 \int_0^\infty \frac{d\mu^2}{(\mu^2+k^2)^2}\at(\mu^2)\;.
\eeq
In the perturbative domain $\as\ll 1$ the standard and effective
couplings are approximately the same:
\beq\label{atexpn}
 \at(\mu^2) = \as(\mu^2) -\frac{\pi^2}{6}\frac{d^2\as}{d\ln^2\mu^2}
 + \ldots \>=\> \as\,+\, {\cal O}(\as^3)\>.
\eeq
Thus one may regard $\at$ defined by \re{aeffrel} as an 
effective measure of QCD interaction strength, extending the 
physical perturbative coupling down to the non-perturbative domain.

Next we write the strong coupling in the form
\beq\label{coupsplit}
  \as(k^2) \>=\> \aPT(k^2) + \delta\as(k^2)\,,
\eeq
where $\aPT$ is the perturbative coupling and $\delta\as$ is a
modification to the effective interaction strength
at small momentum scales, responsible for non-perturbative effects.
The corresponding  ``effective coupling modification'' $\delta\at$,
which generates the non-perturbative interaction strength
$\delta\as$, must satisfy a dispersion relation of the form
\re{atint}:
\beq\label{alal}
\delta\as(k^2)= k^2\int_0^\infty \frac{d\mu^2}{(\mu^2+k^2)^2} \,
\delta\at(\mu^2)\>.
\eeq
It follows that an arbitrary finite modification of the
effective coupling at low scales would generally
introduce power corrections of the form  $1/k^{2p}$ into the
ultraviolet behaviour of the running coupling $\as$ itself.
As discussed in Ref.~\cite{BPY}, such a modification would
destroy the basis of the operator product
expansion \cite{SVZ}.  One must therefore require
that at least the first few integer moments of the coupling
modification should vanish:
\beq\label{vanish}
 \int_0^\infty \frac{d\mu^2}{\mu^2}
\>\left(\mu^2\right)^p \delta\at(\mu^2)\>=\>0\>; \quad p=1,\ldots,p_{\max}\,. 
\eeq
The upper bound $p_{\max}$ could be set by instanton--anti-instanton
contributions at $p_{\max}< \beta_0\sim 9$.

The effect on some observable $\hat F$ of gluon vacuum
polarization insertions in one-loop graphs is represented
in terms of the spectral function $\rho_s(\mu^2)$ by a 
{\it characteristic function} $\cF(\mu^2)$, as follows:
\beq\label{WDM1}
 \hat F \>=\> \as(0)\cF(0) + \int_0^\infty 
\frac{d\mu^2}{\mu^2}\rho_s(\mu^2)
\cdot\cF(\mu^2)
\>=\> \int_0^\infty \frac{d\mu^2}{\mu^2}\rho_s(\mu^2)
\cdot
\left[\,\cF(\mu^2) - \cF(0)\,\right],
\eeq
where we have made use of the formal relation \re{nonAb-disrel} 
to eliminate $ \as(0)$. The characteristic function is obtained
by computing the relevant graphs with a
non-zero gluon mass $\mu$ \citd{hadro}{BBB}.
Note that we do not intend to imply that the gluon has a
real effective mass, but only that the dispersive
representation \re{nonAb-disrel} can be expressed in this way.  

Introducing the effective coupling $\at(\mu^2)$ using
Eq.~\re{aeffrel} and integrating by parts, we can write
\beq\label{WDMF}
  \hat F(x,Q^2)\>=\>  \int_0^\infty
\frac{d\mu^2}{\mu^2}\> \at(\mu^2) \cdot \dot\cF (x,Q^2;\mu^2) \>,
\;\;\;\;\;\;\;\;
\quad  \dot{\cF} \equiv -\frac{\partial\cF}{\partial\ln \mu^2}
\>.
\eeq
Here we have taken $\hat F$ to represent a quark structure function,
with dependence on the deep inelastic scattering variables $x$
and $Q^2$. To obtain the corresponding (non-singlet) hadron structure
function $F(x,Q^2)$, we have to convolute
the quark structure function with the appropriate combination of
quark distribution functions $q(x)$. As discussed in Ref.~\cite{BPY},
one finds that the characteristic function $\cF$ has
a collinear divergent part, which generates the
scale dependence of the quark distributions and the
usual logarithmic scaling violations. The remaining
part $\Freg$ generates the coefficient function $C$
in the relation between the structure function and the
quark distribution:
\beq
F(x,Q^2) = \int_x^1\frac{dz}{z} C(z,Q^2)\,q(x/z,Q^2)
\eeq
where 
\beq\label{WDM}
 C(x,Q^2)\>=\>  \int_0^\infty
\frac{d\mu^2}{\mu^2}\> \at(\mu^2) \cdot\dot\Freg(x,Q^2;\mu^2) \>.
\eeq

Since $\cF$ depends only on dimensionless ratios, we may write
\beq
\cF (x,Q^2;\mu^2) =\cF (x,\eps)\;,
\;\;\;\;\;\;
\dot{\cF} \equiv -\eps\frac{\partial}{\partial\eps}\cF (x,\eps)\;,
\;\;\;\;\;\;
\eps\equiv\frac{\mu^2}{Q^2}\;.
\eeq
The power-behaved contributions depend on the small-$\eps$
behaviour of $\cF$, which is of the generic form
\beq\label{Fsmalleps}
\cF(x,\eps) = -P(x)\ln\eps + C_0(x) - C_2(x)\eps\ln\eps
- \half C_4(x)\eps^2\ln\eps +\cdots\;.
\eeq
The dots indicate terms that are analytic and vanishing at
$\eps=0$. Thus according to the above definition the
regular part is
\beq\label{Freg}
\Freg(x,\eps) = C_0(x) - C_2(x)\eps\ln\eps
- \half C_4(x)\eps^2\ln\eps +\cdots\;,
\eeq
and
\beq\label{dFreg}
\dot\Freg(x,\eps) = C_2(x)\eps\ln\eps
+C_4(x)\eps^2\ln\eps +\cdots\;.
\eeq

The constraint \re{vanish} means that only those terms
in the small-$\eps$ behaviour of the characteristic function
that are {\em non-analytic} at $\eps =0$ will lead to
power-behaved non-perturbative contributions \cite{BBB}.
According to Eq.~\re{WDM}, the corresponding contributions to
the coefficient function will be of the form
\beq
\CNP(x,Q^2) = \int_0^\infty \frac{d\mu^2}{\mu^2}\, 
\delta\at(\mu^2)\dot\Freg(x,\eps=\mu^2/Q^2)\,. 
\eeq
Thus from the small-$\eps$ behaviour \re{dFreg} we find
\beq\label{CNP}
\CNP(x,Q^2) = C_2(x)\frac{\Apr_2}{Q^2} + C_4(x)\frac{\Apr_4}{Q^4}\;,
\eeq
where, following Ref.~\cite{BPY}, we have defined the log-moment integrals
\beq\label{adefs}
\Apr_{2p} \>=\>\frac{C_F}{2\pi}
\int_0^\infty\frac{d\mu^2}{\mu^2}
\>\mu^{2p}\,\ln(\mu^2/\mu_0^2)\>\,\delta\at(\mu^2)\;.
\eeq
Notice that since integer $\mu^2$-moments of $\delta\at$ vanish,
these quantities are independent of the scale $\mu_0^2$.  For
convenience, we extract a universal factor of $C_F/2\pi$
from the characteristic function.

The interpretation of Eq.~\re{CNP} in terms of a universal
low-energy effective coupling is optional. More generally,
we could interpret this expression as the ambiguity
in the perturbative evaluation of the coefficient function,
arising from the factorial divergence of the perturbation
series generated by gluon vacuum polarization insertions.
In this language, the two terms correspond to infrared
renormalons, and the constants $\Apr_2$ and $\Apr_4$
are proportional to powers of the QCD scale $\Lambda$,
the constants of proportionality depending on how
we choose to resolve the renormalon ambiguity.

\mysection{Calculations}

In this section we apply the dispersive method
to compute the power correction terms \re{CNP}
for non-singlet structure functions.
Recall that the object of central importance is the
characteristic function $\cF(\eps)$ for the emission of a
gluon with mass-squared $\mu^2 = \eps Q^2$ at the hard scale $Q^2$.
For power corrections, the relevant contributions are given by the
non-analytic terms in the small-$\eps$ behaviour of the
logarithmic derivative $\dot\cF(\eps)$.

The characteristic function for the structure function
$F_2$ (actually $F_2/x$) was given in Ref.~\cite{BPY}:
\beq\label{F2t}
\cF_2(x,\eps) = \cFr_2(x,\eps)\,\Theta(1-x-x \eps)
+\cFv(\eps)\,\delta(1-x)\;.
\eeq
The contribution from real gluon emission is
\beq\label{F2r}\eqalign{
\cFr(x,\eps) &= \left[\frac {2(1-\eps)^2}{1-x}-(1+x)
+2(2+x+6x^2)\eps -2(1+x+9x^3)\eps^2\right]
\ln\left[\frac{(1-x \eps)(1-x)}{x^2\eps}\right] 
\cr&
-\frac{3+14\eps-15\eps^2}{2(1-x)} +\frac{\eps}{(1-x)^2}
+\frac{\eps^2}{2(1-x)^3}+\frac{x}{1-x \eps}
\cr& 
+1+3x+6(1-x)(1+3x)\eps-(8+9x+18x^2)\eps^2\;.
}\eeq
The virtual contribution is
\beq\label{Fv}
\cFv(\eps) =
2(1-\eps)^2\left[\Li(\eps)+\ln\eps\ln(1-\eps)
-\frac 12\ln^2\eps -\frac{\pi^2}{3} \right]  
-\frac72 -(3-2\eps)\ln\eps+2\eps 
\,,
\eeq
where 
\beq
\Li(\eps)= -\int_0^\eps\frac{dt}{t}\ln(1-t)\,.
\eeq
We note the relation
\beq\label{Mvint}
\cFv(\eps) = -\int_0^1 \cFr(x;\eps)\Theta(1-x-x \eps)\,dx\;,
\eeq
which means that the Adler sum rule is satisfied identically, that is,
it receives neither perturbative nor power corrections (see Ref.~\cite{AEM}).

Taking the small-$x$ limit, we obtain an expression
of the form \re{Fsmalleps}.
The coefficient of $-\ln \eps $ in $\cFr$ is the quark
splitting function $P(x)=(1+x^2)/(1-x)$, which is singular
for $x\to 1$. The singularity is regularized by including the 
virtual contribution. As discussed above, this term produces the
usual logarithmic scaling violation.
The second term $C_0(x)$ is the perturbative
coefficient function (in the gluon mass regularization
scheme). The remaining terms are analytic at $\eps =0$,
except for two terms proportional to $\eps\ln\eps$ and
$\eps^2\ln\eps$. Thus we obtain only two power corrections,
of the form \re{CNP}.

When taking the $\eps\to 0$ limit of eq.~\re{F2r}, we have to be
careful with the functions that become singular in this limit.
Defining `+', `++' and `+++' prescriptions
such that, for any test function $f$,
\beq\eqalign{
  \int_0^1 F(x)_+\,f(x)\,dx &= \int_0^1 F(x)\,[f(x)-f(1)]\,dx\,\cr
  \int_0^1 F(x)_{++}\,f(x)\,dx &= \int_0^1
F(x)\,[f(x)-f(1)+(1-x)f'(1)]\,dx\,\cr
  \int_0^1 F(x)_{+++}\,f(x)\,dx &= \int_0^1
F(x)\,[f(x)-f(1)+(1-x)f'(1)-\half(1-x)^2 f''(1)]\,dx
}\eeq
and recalling that
\beq
\int_0^1 \delta^{(n)}(1-x)\,f(x)\,dx = f^{(n)}(1)\;,
\eeq
we should replace the singular terms in
Eq.~(\ref{F2r}) at small $\eps$,
up to terms of order $\eps^2$, as follows:
\beq\label{smalleps}\eqalign{
  \frac{1}{1-x} &\to \frac{1}{(1-x)_+}
+(\eps-\half\eps^2-\ln\eps)\,\delta(1-x) \cr
& +\eps(1-\eps)\,\delta'(1-x)-\quart\eps^2\,\delta''(1-x) \cr
  \frac{\ln(1-x)}{1-x} &\to \left(\frac{\ln(1-x)}{1-x}\right)_+
+(\eps\ln\eps-\half\eps^2\ln\eps-\half\ln^2\eps-\half\eps^2)\,\delta(1-x)
\cr
&+\eps(\ln\eps-\eps\ln\eps-1)\,\delta'(1-x)
 -\quart\eps^2(\ln\eps-\half)\,\delta''(1-x)\cr
  \frac{\eps}{(1-x)^2} &\to \frac{\eps}{(1-x)^2_{++}}
+\delta(1-x)+\eps(\ln\eps-\eps)\,\delta'(1-x)
+\half\eps(1-\eps)\,\delta''(1-x)\cr
  \frac{\eps^2}{(1-x)^3} &\to \frac{\eps^2}{(1-x)^3_{+++}}
+(\half+\eps)\,\delta(1-x)-\eps\,\delta'(1-x)
-\half\eps^2\ln\eps\,\delta''(1-x)\,.
}\eeq

Individual terms of the form $\eps^p\ln\eps$ with $p>2$ are
generated by the above replacements, but they cancel in the
sum \re{F2t}. The only remaining terms in $\Freg$ that are
non-analytic as $\eps\to 0$ are those of the form $\eps\ln\eps$
and $\eps^2\ln\eps$.

The characteristic function for the structure function $2F_1$
is given by
\beq
\cF_1(x,\eps) = \cF_2(x,\eps) - \cF_L(x,\eps)
\eeq
where $\cF_L$, the characteristic function for the
longitudinal contribution $F_L/x$, is
\beq\label{FL}\eqalign{
\cF_L(x,\eps) &= 4(2-3x\eps)x^2\eps
\ln\left[\frac{(1-x \eps)(1-x)}{x^2\eps}\right] 
-\frac{2\eps(2-\eps)}{1-x} +\frac{2\eps^2}{(1-x)^2}
\cr&
+2x+4(1-x)(1+3x)\eps-2(2+3x+6x^2)\eps^2\;.
}\eeq
This contribution introduces no new power corrections,
as noted in Ref.~\cite{stein}.

The characteristic function for the parity-violating
structure function $F_3$ was given in Ref.~\cite{BPY}:
\beq
\cF_3(x,\eps) = \cF_2(x,\eps) - \cF_d(x,\eps)
\eeq
where
\beq\label{Fd}\eqalign{
\cF_d(x,\eps) &= 2(4+\eps-9x\eps)x^2\eps
\ln\left[\frac{(1-x \eps)(1-x)}{x^2\eps}\right] 
-\frac{\eps(4-3\eps)}{1-x} +\frac{2\eps^2}{(1-x)^2}
+\frac{2x}{1-x\eps}
\cr&
+1-x+2(2+5x-9x^2)\eps-(5+7x+18x^2)\eps^2\;.
}\eeq
Again, no power corrections beyond those in Eq.~\re{CNP}
are introduced by this contribution. 

The power corrections to the coefficient function for the
polarized structure function $g_1$ are the same as those
for $F_3$, since their characteristic functions are identical
to one-loop order.

\mysection{Results}

The coefficients in Eq.~\re{CNP} for the power
corrections to the coefficient function for $F_2/x$
are found from Eqs.~\re{F2t}-\re{smalleps} to be
\beq\eqalign{
 C_2(x) &= -\frac{4}{(1-x)_+} +2(2+x+6x^2)
-9\,\delta(1-x)-\delta'(1-x) \cr
 C_4(x) &=  \frac{4}{(1-x)_+} -4(1+x+9x^3)
+15\,\delta(1-x)+\frac{1}{2}\,\delta''(1-x)\;.
}\eeq
The corresponding expressions in moment space,
defined by
\beq
\tilde C(N) = \int_0^1 x^{N-1}\,C(x)\,dx
\eeq
are
\beq\eqalign{
\tilde C_2(N) &= -N -8 +\frac 4N +\frac{2}{N+1}+\frac{12}{N+2}+4S_1 \cr
\tilde C_4(N) &= \frac 12 N^2 -\frac 32 N +16 -\frac 4N -\frac{4}{N+1}
-\frac{36}{N+3}-4S_1\;,
}\eeq
with
\beq\label{Sdefs}
  S_1 = \sum_{j=1}^{N-1} \frac 1j = \psi(N)+\gamma_E = \ln N+{\cal O}(1/N)\,.
\eeq

For $2F_1=F_2/x-F_L/x$, the corresponding results are\footnote{Our
results for $F_L$ agree with those of Ref.~\cite{stein}.}
\beq\eqalign{
 C_2(x) &= -\frac{4}{(1-x)_+} +2(2+x+2x^2)
-5\,\delta(1-x)-\delta'(1-x) \cr
 C_4(x) &= \frac{4}{(1-x)_+} -4(1+x+3x^3)
+11\,\delta(1-x)+4\,\delta'(1-x)+\frac{1}{2}\,\delta''(1-x)\,,
}\eeq
\beq\eqalign{
\tilde C_2(N) &= -N -4 +\frac 4N +\frac{2}{N+1}+\frac{4}{N+2}+4S_1 \cr
\tilde C_4(N) &= \frac 12 N^2 +\frac 52 N +8 -\frac 4N -\frac{4}{N+1}
-\frac{12}{N+3}-4S_1\;.
}\eeq

For $F_3$ (and $g_1$),
\beq\eqalign{
 C_2(x) &= -\frac{4}{(1-x)_+} +2(2+x+2x^2)
-5\,\delta(1-x)-\delta'(1-x) \cr
 C_4(x) &= \frac{4}{(1-x)_+} -4(1+x+x^2)
+9\,\delta(1-x)+4\,\delta'(1-x)+\frac{1}{2}\,\delta''(1-x)\,,
}\eeq
\beq\eqalign{
\tilde C_2(N) &= -N -4 +\frac 4N +\frac{2}{N+1}+\frac{4}{N+2}+4S_1 \cr
\tilde C_4(N) &= \frac 12 N^2 +\frac 52 N +6 -\frac 4N -\frac{4}{N+1}
-\frac{4}{N+2}-4S_1\;.
}\eeq
Note that the $1/Q^2$ coefficients $C_2(x)$ are the same for $F_3$
and $2F_1$, but the $1/Q^4$ coefficients $C_4(x)$ are slightly different.

\mysection{Discussion}

To illustrate the above results we examine the $1/Q^2$ and $1/Q^4$
contributions arising from the valence quark distributions in neutrino
scattering.  Defining $F = \half(F^{\nu}+F^{\bar\nu})_V$ and
$q = u_V + d_V$, where $V$ indicates the valence contribution,
we can write
\beq\label{Ftwist}\eqalign{
F(x,Q^2) &\simeq q(x,Q^2)\left(1+\frac{D_2(x,Q^2)}{Q^2}
+\frac{D_4(x,Q^2)}{Q^4}\right)\cr 
D_{2p}(x,Q^2) &= \frac{\Apr_{2p}}{q(x,Q^2)}
\int_x^1\frac{dz}{z} C_{2p}(z)\,q(x/z,Q^2)\;,
}\eeq
the coefficient functions $C_{2p}$ being as given above
for $F=F_2/x$, $2F_1$ and $F_3$. The valence quark
distributions were taken from the MRSA parametrization \cite{MRSA}.
Calculations were performed at $Q^2=10$ GeV$^2$, but the
$Q^2$-dependence of the coefficients  $D_{2p}(x,Q^2)$ is negligible.

\begin{figure}
\vspace{9.0cm}
\includegraphics{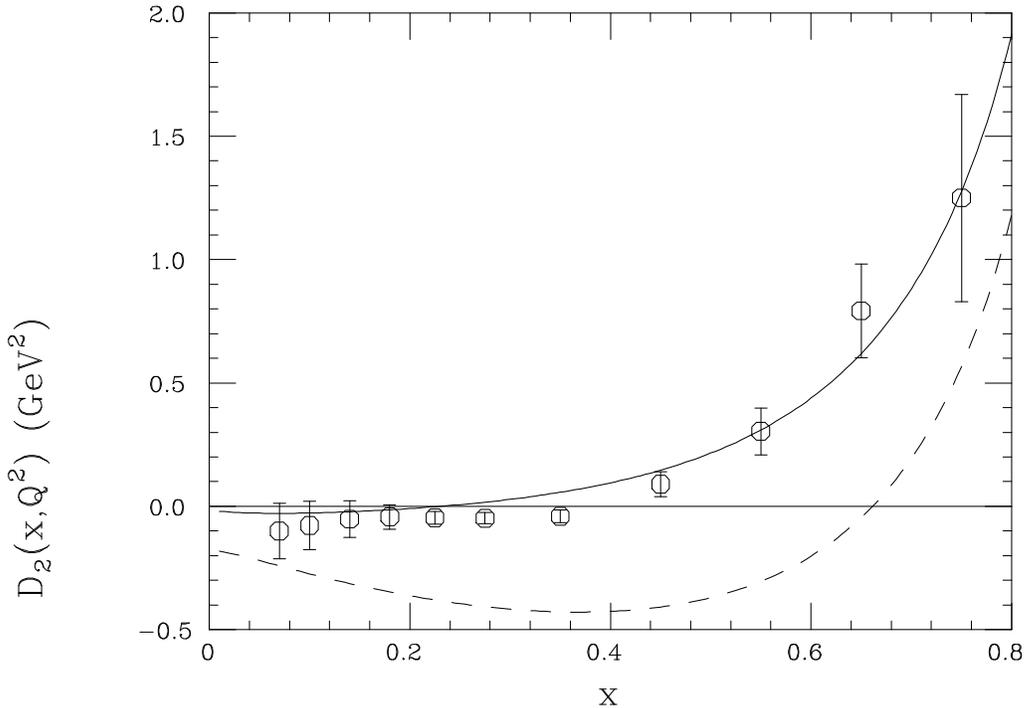}
\caption{Coefficients of $1/Q^2$ contributions to $F_2$ (solid),
and to $F_1$ and $F_3$ (dashed).}
\label{fig_twist4}
\end{figure}

Figure~\ref{fig_twist4} shows the $1/Q^2$ coefficients $D_2(x,Q^2)$,
assuming the value $\Apr_2 = -0.2$ GeV$^2$ for the non-perturbative
parameter defined by Eq.~\re{adefs} with
$p=1$. We see that this gives remarkably good agreement with the
data points for $F_2$, taken from Ref.~\cite{VM}.\footnote{The
data are for charged leptons but the predicted coefficients
$D_{2p}(x,Q^2)$ for neutral and charged lepton scattering
are essentially identical.}
The generally negative value of the
prediction for $F_3$ leads to a negative
correction to the Gross--Llewellyn-Smith sum rule, in
qualitative agreement with the predictions of Ref.~\cite{BK},
used in the test of the sum rule by the CCFR-NuTeV
Collaboration \cite{harris}.

\begin{figure}
\vspace{9.0cm}
\includegraphics{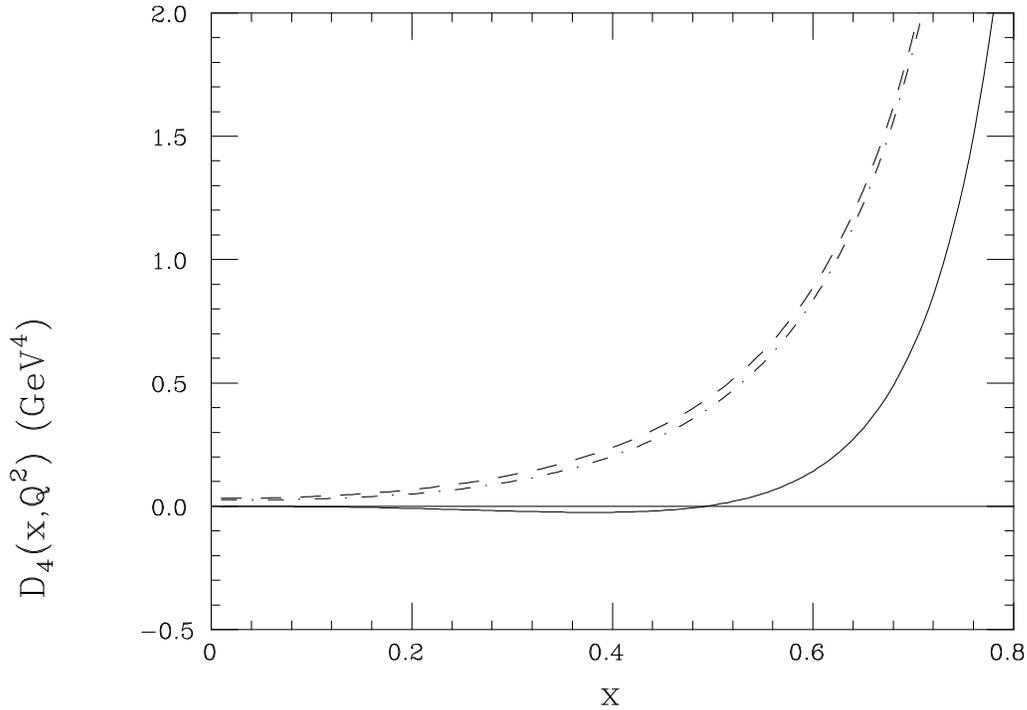}
\caption{Coefficients of $1/Q^4$ contributions to $F_2$ (solid),
$F_1$ (dashed) and $F_3$ (dot-dashed).}
\label{fig_twist6}
\end{figure}

Figure~\ref{fig_twist6} shows corresponding predictions
for the $1/Q^4$ coefficients $D_4(x,Q^2)$.
Here the arbitrary choice
$\Apr_4 = (\Apr_2)^2 = 0.04$ GeV$^4$ is made
for illustration only. We see that the $1/Q^4$
contributions are peaked more sharply at high $x$,
owing to the $\delta''(1-x)$ term in $C_4(x)$. This
indicates that the power corrections are functions
of $(1-x)Q^2$ rather than $Q^2$ at high $x$.

It would clearly be of interest to attempt a global fit to deep inelastic
data at moderate $Q^2$, treating the quantities $\Apr_2$ and $\Apr_4$
as free parameters. For this purpose, a program to compute the predicted
power corrections is available from the authors.

\par \vskip 1ex
\noindent{\large\bf References}
\begin{enumerate}
\item\label{BroKat}
      D.J.\ Broadhurst and A.L.\ Kataev, \pl{315}{179}{93}.
\item\label{ji}
      X.-D.\ Ji, \np{448}{51}{95}. 
\item\label{lomax}
     C.N.\  Lovett-Turner and C.J.\ Maxwell, \np{452}{188}{95}. 
\item\label{stein}
       E.\ Stein, M.\ Meyer-Hermann, L.\ Mankiewicz and A.\ Sch\"afer,
       Frankfurt preprint UFTP 407/1996 [hep-ph/9601356].
\item\label{BPY}
       Yu.L.\ Dokshitzer, G.\ Marchesini and B.R.\ Webber, CERN
       preprint CERN-TH/95-281 [hep-ph/9512336], to be published
       in Nucl.\ Phys.\ B.
\item\label{renormalons}
       For reviews and classic references see
       V.I. Zakharov, \np{385}{452}{92} and 
       A.H.\ Mueller, in {\em QCD 20 Years Later}, vol.~1
       (World Scientific, Singapore, 1993).
\item\label{SVZ}
    M.A. Shifman, A.I. Vainstein and V.I. Zakharov, 
    \np{147}{385,448,519}{79};\\
       {\it Vacuum Structure and QCD Sum Rules: Reprints},
       ed.\ M.A.\ Shifman (North-Holland, 1992:
       Current Physics, Sources and Comments, v.\ 10).
\item\label{hadro}
       B.R.\ Webber, \pl{339}{148}{94}.
\item\label{BBB}
   M. Beneke, V.M. Braun and V.I. Zakharov, \prl{73}{3058}{94};\\
   P.\ Ball, M.\ Beneke and V.M.\ Braun, \np{452}{563}{95};\\
   M.\ Beneke and V.M.\ Braun, \np{454}{253}{95}.
\item\label{AEM}
       G. Altarelli, R.K. Ellis and G. Martinelli, 
\np{143}{521}{78};\\
       J. Kubar-Andr\'e and F.E. Paige, \pr{19}{221}{79}.
\item\label{MRSA}
       A.D. Martin, R.G. Roberts and W.J. Stirling, \pr{50}{6734}{94}.
\item\label{VM}
       M.\ Virchaux and A.\ Milsztajn, \pl{274}{221}{92}.
\item\label{BK}
     V.M.\ Braun and A.V.\ Kolesnichenko, \np{283}{723}{87};
\item\label{harris}
     D.A.\ Harris, CCFR-NuTeV Collaboration, FERMILAB-CONF-95-144,
     to appear in {\em Proc.\ 30th Rencontres de Moriond, Meribel les
     Allues, France, March 1995}. 
\end{enumerate}
\end{document}